\documentclass{webofc}
\usepackage[varg]{txfonts}   % Web of Conferences font
\begin{document}
\title{Towards precision SZ cluster cosmology: from \textit{Planck} to the Simons Observatory }

\author{\lastname{\'I. Zubeldia}\inst{1,2}\fnsep\thanks{e-mail: inigo.zubedia@ast.cam.ac.uk}
        % etc.
}

\institute{Institute of Astronomy, University of Cambridge, Madingley Road, Cambridge CB3 0HA
\and
         Kavli Institute for Cosmology, University of Cambridge, Madingley Road, Cambridge CB3 0HA          }

\abstract{
As demonstrated by \textit{Planck}, SPT, and ACT, the abundance of Sunyaev-Zeldovich-detected galaxy clusters across mass and redshift is a powerful cosmological probe. Upcoming experiments such as the Simons Observatory (SO) will detect over an order of magnitude more objects than what previous experiments have found, thereby providing an unprecedented constraining potential. However, in order for this potential to be realised, the cluster detection and analysis pipelines will have to be built and understood to a much higher level of accuracy than has been demonstrated to date.

Here we discuss ongoing efforts towards the accurate modelling of tSZ cluster counts, focusing on the improvements regarding optimisation bias, covariance estimation, and foreground deprojection of \cite{Zubeldia2021,Zubeldia2022a,Zubeldia2022b}, which are implemented in the publicly-available \texttt{SZiFi} package. Next, we briefly discuss the application of these improved cluster detection methods to \textit{Planck} data. Finally, we introduce \texttt{cosmocnc}, a new cluster number count likelihood code that will be publicly available soon.
}
\maketitle
\section{Introduction}
\label{intro}

The abundance of galaxy clusters as a function of mass and redshift has long been recognised as a powerful cosmological probe, sensitive to cosmological parameters such as $\Omega_{\mathrm{m}}$, $\sigma_8$ and the equation of state of dark energy (see, e.g., \cite{Allen}). The thermal Sunyaev-Zeldovich (tSZ) effect \cite{Sunyaev} offers a unique window into the cluster population, allowing for cluster detection to high redshift. These tSZ-selected catalogues can be, in turn, used for cosmological inference, as demonstrated by \textit{Planck}, SPT, and ACT (see, e.g., \cite{Planck,Bleem,Hilton,Bocquet,Zubeldia2019}). Upcoming mm experiments such as the Simons Observatory and CMB-S4 are set to revolutionise cluster cosmology, with them expected to find about 20\,000 and $10^5$ clusters, respectively \cite{Simons,S4}. These numbers will come with an unprecedented constraining potential. However, in order for this potential to be realised, the analysis pipeline, from cluster detection to cosmological parameter inference, will have to be constructed and understood to a much higher level of accuracy than what has been demonstrated in previous analyses.

In this contribution, we outline several several analysis improvements made towards the maximal exploitation of upcoming tSZ cluster cosmology data. In particular, in Section\,\ref{sec-1} we describe two significant improvements to the multi-frequency matched filter (MMF) cluster detection method, namely iterative noise covariance estimation and foreground spectral deprojection, which have been implemented in the \texttt{SZiFi} package. Next, in Section\,\ref{sec-2} we briefly discuss the application of these enhanced cluster detections methods to \textit{Planck} data, in what is a currently ongoing project. Finally, in Section\,\ref{sec-3} we introduce \texttt{cosmocnc}, a soon-to-be-released cluster-number-count theory package designed for fast likelihood computation.

\section{\texttt{SZiFi}: a new MMF cluster finder with iterative noise covariance estimation and foreground deprojection}
\label{sec-1}

Multi-frequency matched filters (MMFs) \cite{Melin} have become the standard tool with which clusters are detected at mm wavelengths. They rely on our knowledge of the cluster tSZ signal, both spectrally, as given by the tSZ spectral energy distribution (SED), and spatially, as given by the cluster pressure profile. Here we introduce \texttt{SZiFi}, the Sunyaev-Zeldovich iterative Finder, a new implementation of the MMF cluster finding method that has been developed in order to study systematics in cluster detection and, ultimately, to apply it to upcoming mm data. \texttt{SZiFi} is fully written in Python and publicly available\footnote{\texttt{github.com/inigozubeldia/szifi}}. It incorporates a number of novel features, most notably iterative noise covariance estimation and foreground spectral deprojection, which we describe next.

\subsection{Iterative noise covariance estimation}

The motivation for iterative noise covariance estimation is the following. In order for it to be applied, a MMF needs as input the covariance matrix of the noise in the data, where by noise we mean all of the components in the data other than the one that is being targeted, i.e., the cluster tSZ signal. The noise covariance is typically estimated from the data and taken to be equal to the covariance of the data (e.g., \cite{Melin}). However, as we discuss in detail in \cite{Zubeldia2022a}, doing this creates two problems. First, the noise covariance is overestimated, which leads to a loss of signal-to-noise. Second, as the covariance is estimated from the data, the response of the MMF to the data becomes nonlinear. As we show in \cite{Zubeldia2022a}, if the tSZ signal is present in the data-estimated covariance, a bias is induced in the cluster observables (cluster Compton-$y$ estimate and signal-to-noise). This bias, which is analogous to the ILC bias encountered in map-based component separation (see, e.g., \cite{Coulton}), can be as large as 0.5\,$\sigma$ \emph{per cluster}, and therefore potentially significant in a cosmological analysis.

\begin{figure}[h]
\centering
\includegraphics[scale=0.6]{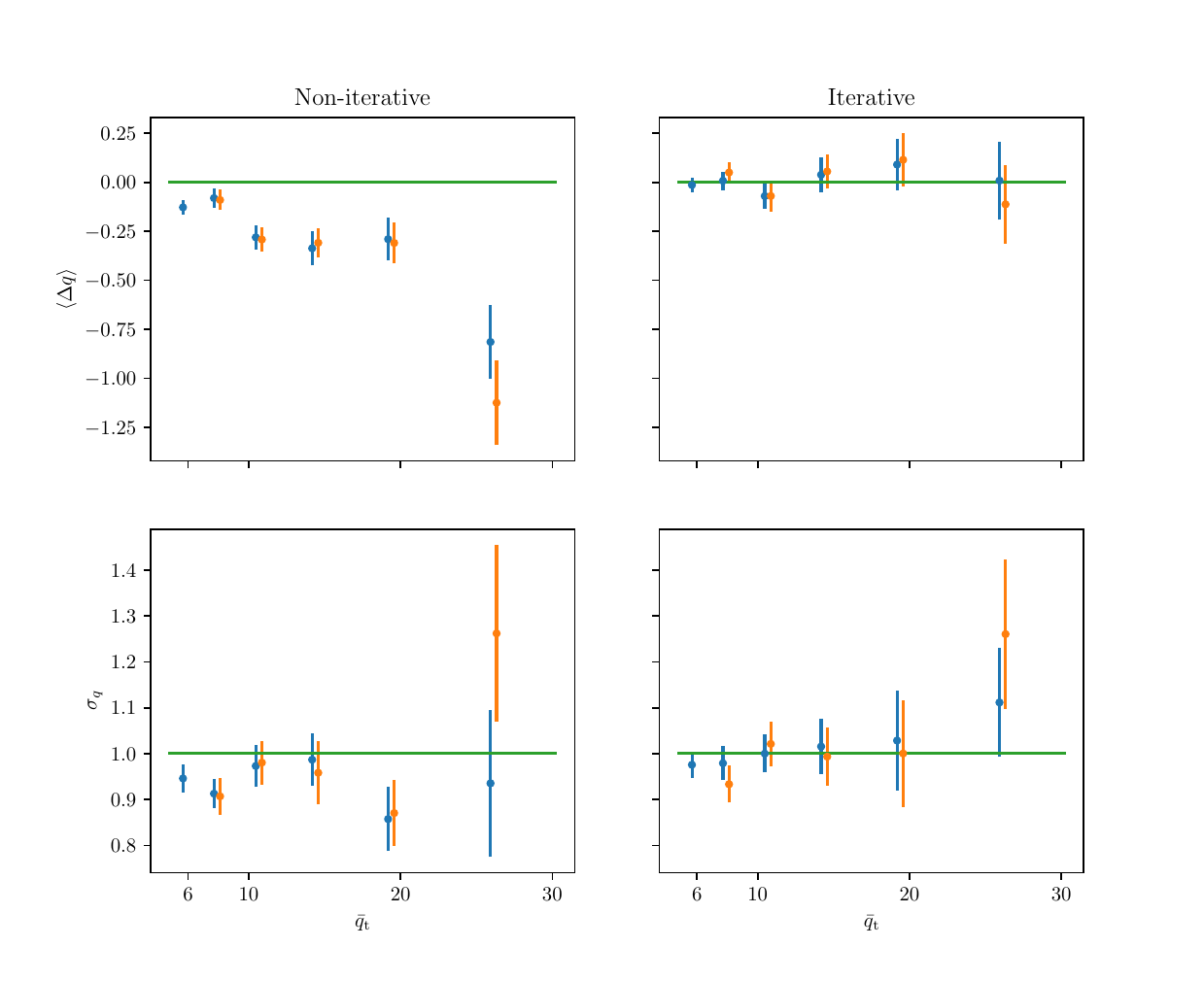}
\caption{Empirical mean and standard deviation (upper and lower panels, respectively) of the signal-to-noise measurement residuals (difference between measured and mean signal-to-noise) of the simulated \textit{Planck}-like catalogues analysed in \cite{Zubeldia2022a}, binned in true signal-to-noise for both non-iterative (left panels) and iterative (right panel) covariance
estimation. The blue points correspond to the ‘fixed’ case, in which the signal-to-noise is extracted at the true cluster parameters, whereas the orange points correspond to the optimal signal-to-noise, obtained blindly by maximisation over angular scale and sky location. A bias in both the mean of the signal-to-noise and on its standard deviation is observed in the non-iterative case. On the other hand, these two statistics are consistent with no bias and unit value, respectively, in the iterative case, demonstrating the power of iterative noise covariance estimation. Figure taken from \cite{Zubeldia2022a}.}
\label{fig-1}       % Give a unique label
\end{figure}

As demonstrated in \cite{Zubeldia2022a}, iterative noise covariance estimation is a highly effective solution for these two problems. In this approach, the noise covariance is first estimated by taking it to be equal to the data covariance, which leads to a first, non-iterative cluster catalogue. Significant detections are then masked from the data and the noise covariance is re-estimated and used in a second run of the MMF algorithm, which delivers an updated, iterative cluster catalogue. If the masking signal-to-noise threshold is low enough, one iteration suffices to completely remove the ILC-like bias and to boost the signal-to-noise of the detections to the expected level (see Fig.~\ref{fig-1}).

\subsection{Foreground spectral deprojection}

As we argue in \cite{Zubeldia2022b}, any signal that is spatially correlated with the tSZ field will induce a bias in the MMF cluster observables (cluster Compton-$y$ estimate and signal-to-noise). In particular, simulations suggest that the Cosmic Infrared Background (CIB) can cause significant biases \cite{Zubeldia2022b}. In order to address this problem, we have developed a spectrally constrained MMF that is able to completely null, or `deproject', the contribution from one or several foregrounds with given SEDs at the expense of a certain signal-to-noise penalty.

\begin{figure}[h]
\centering
\includegraphics[scale=0.25]{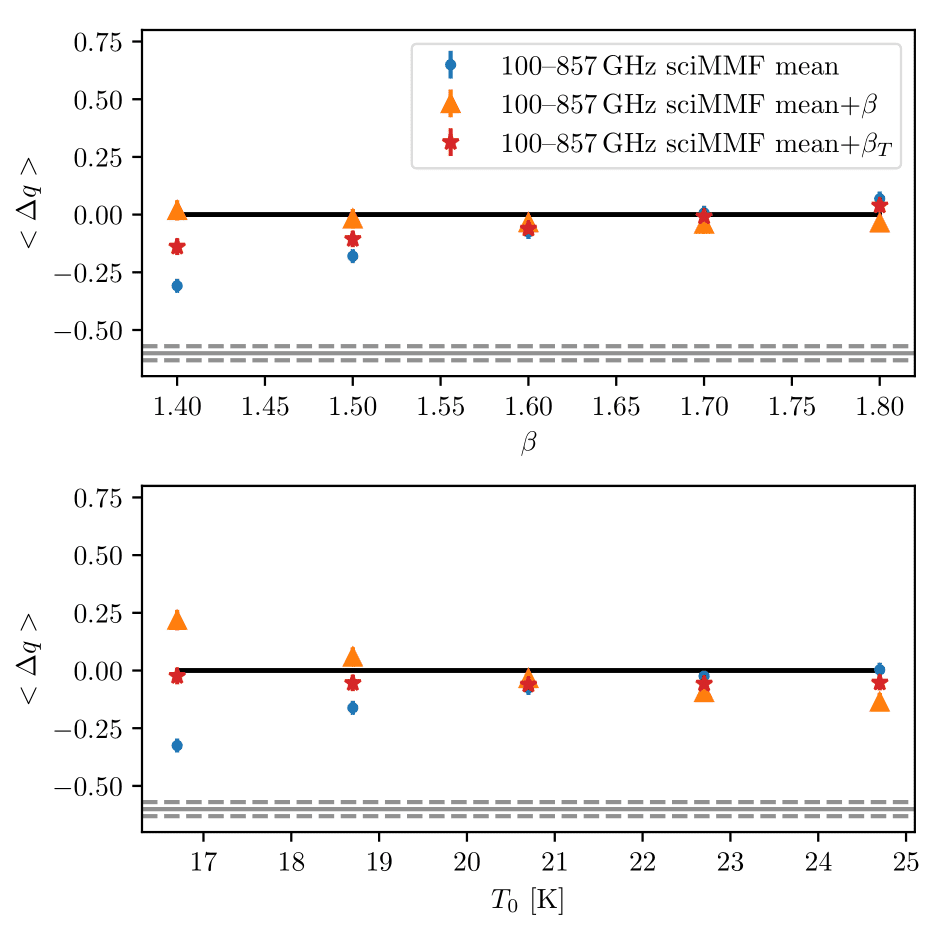}
\caption{Empirical mean of the signal-to-noise residuals for three of the catalogues analysed in \cite{Zubeldia2022b}, constructed by applying spectrally constrained iterative MMFs, or sciMMFs, to simulated \textit{Planck}-like data. In particular, the blue, orange, and red data points correspond, respectively, to catalogues obtained by deprojecting a fiducial CIB SED (blue) and the same fiducial CIB SED in addition to its first-order moment with respect to $\beta$ and $\beta_T$ (orange and red, respectively). This is shown as a function of the values of $\beta$ and $T_0$ (dust temperature at $z=0$) at which the CIB SED that is deprojected is evaluated. The central values of both panels correspond to the true, input values of $\beta$ and $T_0$, while the grey lines show the bias in the standard iMMF case, i.e., with no deprojection. This plot clearly illustrates the effectiveness of our sciMMFs at suppressing the CIB-induced bias in the cluster signal-to-noise.}
\label{fig-2}       % Give a unique label
\end{figure}

In the particular case of the CIB, its SED is not perfectly well constrained and, even if it were, it is not fixed but varies as a function of redshift. Due to this, deprojecting some fiducial CIB SED may not be enough to effectively remove the CIB-induced bias from the cluster observables. Following the moment expansion approach of \cite{Chluba}, however, one can also deproject the first-order moments of the CIB SED with respect to some of the parameters describing it (e.g., the spectral emissivity index $\beta$ or the inverse dust temperature at $z=0$, $\beta_T$). As we show in \cite{Zubeldia2022b} for simulated data, our spectrally constrained MMF together with the moment expansion approach can be highly effective at suppressing the CIB-induced bias in the cluster observables (see Fig.~\ref{fig-2}).

\subsection{Optimisation bias}

The signal-to-noise is the preferred cluster observable in tSZ cluster analyses, and indeed is the observable through which the cluster sample is typically selected, with cluster candidates being identified as the peaks in the MMF signal-to-noise maps. This means that the signal-to-noise for each cluster is obtained by maximising the MMF signal-to-noise over a number of parameters, typically three: two sky coordinates and the cluster angular size. In modelling these `optimal' signal-to-noise measurements, the optimisation procedure must be taken into account. As first proposed in \cite{Vanderlinde} and extensively studied in \cite{Zubeldia2021}, this can be done through a very simple analytical prescription accounting for the number of degrees of freedom over which the signal-to-noise is maximised. As we show in \cite{Zubeldia2022a}, this prescription works to a high level of accuracy for the cluster catalogues delivered by \texttt{SZiFi}.

\section{Application to \textit{Planck} data}
\label{sec-2}

As a demonstration of our improved cluster detections methods, we are currently applying them, as implemented in \texttt{SZiFi}, to data from the \textit{Planck} satellite, with the goals of (1) producing new cluster catalogues and (2) obtaining cosmological constraints from them. The catalogues and derived constraints will be published in the coming months.

\section{\texttt{cosmocnc}: a fast and flexible cluster number count likelihood package}
\label{sec-3}

We have developed a new cluster number count likelihood code, \texttt{cosmocnc}, which will be made public soon along with an accompanying paper (Zubeldia \& Bolliet, in prep.). Written in Python, \texttt{cosmocnc} is fast and very flexible, having been designed with the hope that it can be used to in order to perform a cosmological analysis with any cluster sample with little modification. Its main features are the following:

\begin{itemize}

    \item It supports three types of likelihoods: an unbinned likelihood, a binned likelihood, and an extreme value likelihood.
    \item It also supports stacked data (e.g., stacked lensing profiles), which is modelled in a consistent way with the cluster catalogue.
    \item It links the cluster mass observables (e.g., tSZ signal-to-noise, lensing mass estimate, or X-ray luminosity) to the cluster mass and redshift through a hierarchical model with an arbitrary number of layers, allowing for correlated scatter between the different mass observables. In each layer, the mass--observable scaling relations and the scatter covariance matrix can be defined in a custom way and can depend on sky location.
    \item It incorporates several widely-used halo mass functions.
    \item The unbinned likelihood supports an arbitrary number of cluster mass observables for each cluster in the sample, and it allows for the set of mass observables to vary from cluster to cluster. It also allows for redshift measurement uncertainties.
    \item It allows for the presence of non-validated (i.e., potentially false) detections in the catalogue, modelling them in a consistent way.
    \item It also allows for the generation of synthetic cluster catalogues for a given observational set-up, which can be used for, e.g., accuracy tests.
    \item It can parallelise several of its computations using Python's \texttt{multiprocessing} module, boosting its performance.
    \item It is interfaced with the Markov chain Monte Carlo (MCMC) code \texttt{Cobaya} \cite{Torrado}, allowing for easy-to-run MCMC parameter estimation.
\end{itemize}

\texttt{cosmocnc} has been benchmarked against \texttt{class\_sz}\footnote{\texttt{github.com/CLASS-SZ/class\_sz}}, which also allows for cluster number count likelihood calculation, although in more restrictive scenarios, finding excellent agreement between the two codes.

%
% BibTeX or Biber users please use (the style is already called in the class, ensure that the "woc.bst" style is in your local directory)
% \bibliography{name or your bibliography database}

\end{document}